# Pluto Geologic Map: Use of Crater Data to Understand Age Relationships

**K. N. Singer[1], O. L. White[2], S. Greenstreet[3,4], J. M. Moore[5], D. A. Williams[6], and R. M. C. Lopes[7]**

[1] Southwest Research Institute, Boulder, Colorado, USA. [2] Carl Sagan Center at the SETI Institute, Mountain View, California, USA. [3] Rubin Observatory, NSF NOIRLab, Tucson, AZ 85719, USA. [4] Department of Astronomy and the DiRAC Institute, University of Washington, Seattle, Washington, USA. [5] NASA Ames Research Center, Moffett Field, California, USA. [6] School of Earth & Space Exploration, Arizona State University, Tempe, Arizona, USA. [7] Jet Propulsion Laboratory, California Institute of Technology, Pasadena, CA, USA.

Corresponding author: Kelsi N. Singer (kelsi.singer@swri.org)

**Key Points:**

- We developed crater analysis techniques with specific applications for determining stratigraphic sequences on USGS geologic maps.
- Pluto has a wide variety of both younger and older terrain types, implying ongoing activity and a diverse geologic history.
- Young, crater-free, areas with widely varying geomorphology make up a large fraction, about 30%, of Pluto's near-side.

## Abstract

Pluto's surface displays a wide variety of geologic units from smooth plains to extremely rugged mountainous expanses. These terrains range in age from young, actively resurfaced regions (no observable craters even in the highest-resolution New Horizons images) to old, heavily cratered, eroded regions. Here we expand upon the crater data analysis and the independent crater data set used in the production of a 1:7M scale geologic map of Pluto that is to be published by the United States Geologic Survey (USGS). We present both relative ages based on crater spatial density (number of craters in a given size bin per $km^2$) and also quantitative ages (e.g., 2 Ga) using the estimated impactor flux onto Pluto. The techniques presented here were developed specifically for the information available from a USGS geologic map, where smaller craters are mapped as points only (no specific diameter information per crater). We developed a new type of visualization, called a distributed R-plot, to understand the relative ages of the geologic units. The uncertainties in the current knowledge of the Kuiper belt populations and impactor flux at Pluto propagate to large uncertainties in the estimated quantitative ages (~a factor of two). However, both relative and quantitative ages from crater analysis were still valuable tools in developing the sequence of geologic events. Pluto has large areas of crater-free young terrains (13 units making up ~27% of mapped higher-resolution surface area), with widely varying morphologies, indicating a variety of resurfacing mechanisms, both exogenic and endogenic, likely active into Pluto's recent past or present.

## Plain Language Summary

Geologic maps highlight different types of units across a body and can be useful for observing patterns and understanding the geologic history of a body. Craters form continuously across Pluto from the bombardment of other smaller Kuiper belt objects. Thus, a surface with fewer craters has likely been resurfaced. The resurfacing event could partially or completely erase craters, could happen at a more distinct point in time, or be a long-term process. Each of these scenarios leaves different geological clues, both in the number of craters and their erosion states. A large fraction (~30%) of the surface of Pluto observed in detail by the New Horizons spacecraft (Plutos "near-side") is completely devoid of craters, indicating these are very young surfaces and implying relatively recent geologic activity on Pluto. Additionally, the impact rates on Pluto have been estimated and we can get a rough idea of when in the Solar System's ~4.5 billion year history the surfaces on Pluto likely formed.

## 1 Introduction

The New Horizons spacecraft flew through the Pluto system on July 14, 2015, and collected data with 7 instruments, including optical, color, UV, and IR spectral imaging along with particle and plasma data (Stern et al., 2015; Bagenal et al., 2016; Gladstone et al., 2016; Grundy et al., 2016; Moore et al., 2016; Weaver et al., 2016). Pluto was revealed to be a world of remarkably diverse geologic terrains, with both familiar features like craters and extensional tectonics, and terrains completely unique to Pluto such as a giant convecting volatile ice sheet and distinctive undulating terrains that are likely a form of icy volcanism (e.g., Moore et al., 2016; White et al., 2021; Singer et al., 2022).

Co-author Oliver White led the production of a 1:7M scale geologic map of Pluto (White et al., 2023) using a variety of datasets from the New Horizons encounter to help better understand Pluto's complex geologic history (**Figure 1**). This map passed USGS technical review in October 2023 and is in press, expected to be published by the USGS as a Science Investigations Map towards the end of 2024 or early 2025. Space on the map sheet and in the map text was limited, and thus this separate paper was needed to fully describe and expand on the crater analysis methods and results. Because the map is not yet publicly available, we provide the basic map information, summarize the relevant mapping methods, and provide the files relevant to this paper as supplementary information here. The correlation of map units (COMU) (**Figure 2**), shows the units grouped by morphologic type and also displays their estimated temporal relationships to each other and Pluto's overall geologic history.

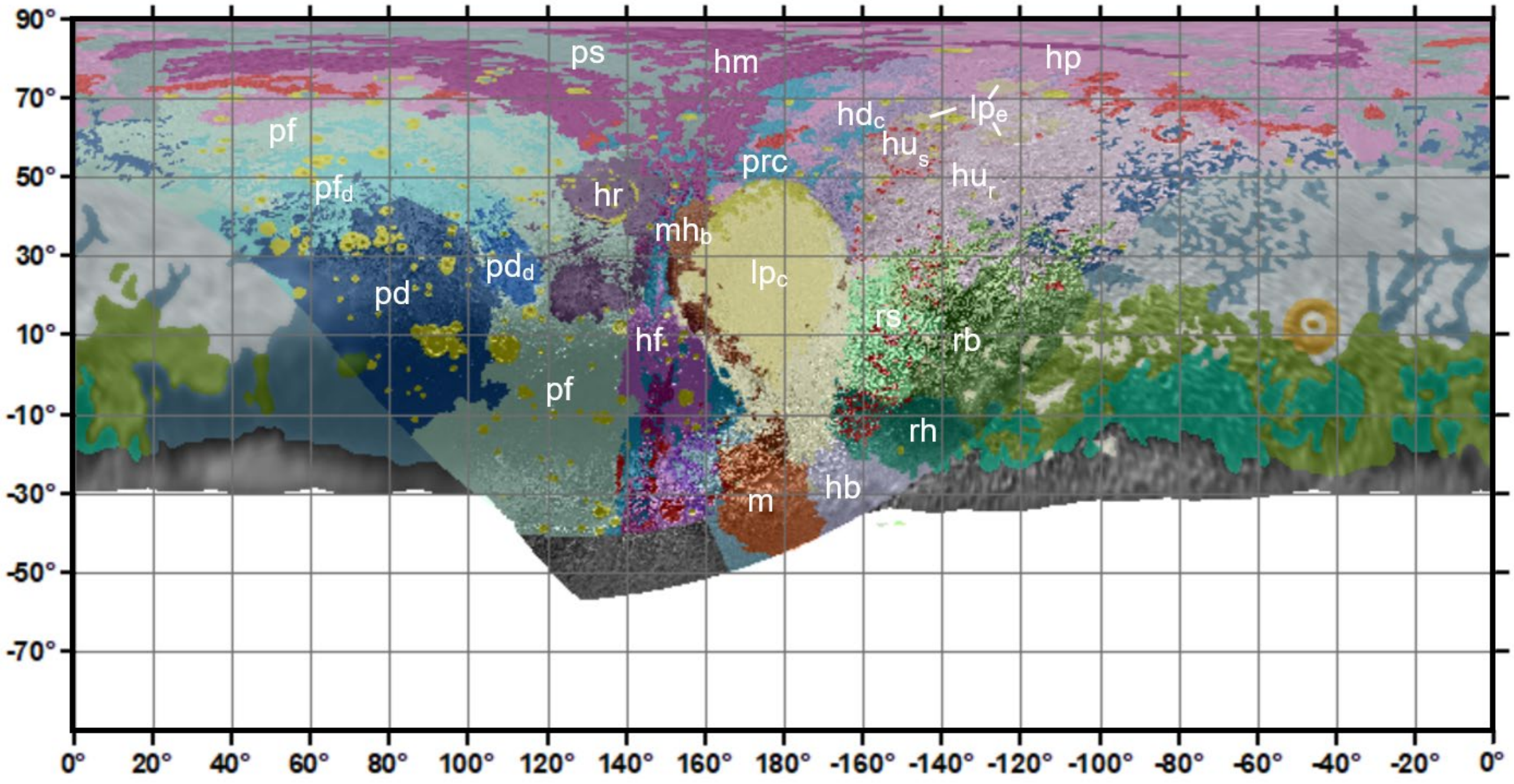

**Figure 1. Overview of the Pluto geologic map and major units across Pluto's near-side.** Simple cylindrical projection. The terrain south of -30° latitude was mostly in darkness during the New Horizons encounter, except for the spur between ~120° and -160° longitude, which was illuminated by light reflected onto the nightside surface by hazes in Pluto's atmosphere.

Every object in the Solar System is unique in terms of both its geology and the datasets available. Thus, different analyses may be appropriate for different locations/maps. Some bodies have too few craters to glean meaningful crater statistics per unit at the given map scale, such as Io (Williams et al., 2011), Europa (Leonard et al., 2024), or Titan (Lopes et al., 2020; Williams et al., 2023).

Two USGS geologic maps preceding the Pluto map contained specific crater analyses per unit used to understand the unit ages: the Tanaka et al. (2014b) geologic map of Mars and the Collins et al. (2013) geologic map of Ganymede. We provide summaries of the crater analyses in these two maps below as context for this current work. Many other works have used craters for age analysis, but we focus only on USGS geologic maps here.

The Tanaka et al. (2014a; 2014b) map of Mars contains extensive crater density analyses. They conducted detailed crater mapping for 48 representative areas from 23 units (out of a total of 44 units), and noted that crater analysis had been conducted by other authors previously for additional units. This data was used for relative age correlations and also for calculating age estimates based on the Ivanov (2001) chronology function (Tanaka et al., 2014b, their Table 2). Separately, the authors pulled out the craters superposing every unit from the Robbins & Hynek (2012) database, and cumulative crater densities per unit for 4 diameter subsets (N[1], N[2], N[5], N[16]) were provided. Additionally, the authors looked at subsets of the crater database based on crater degradation or ejecta type, both globally and for period groups (Tanaka et al., 2014a, their Table 5).

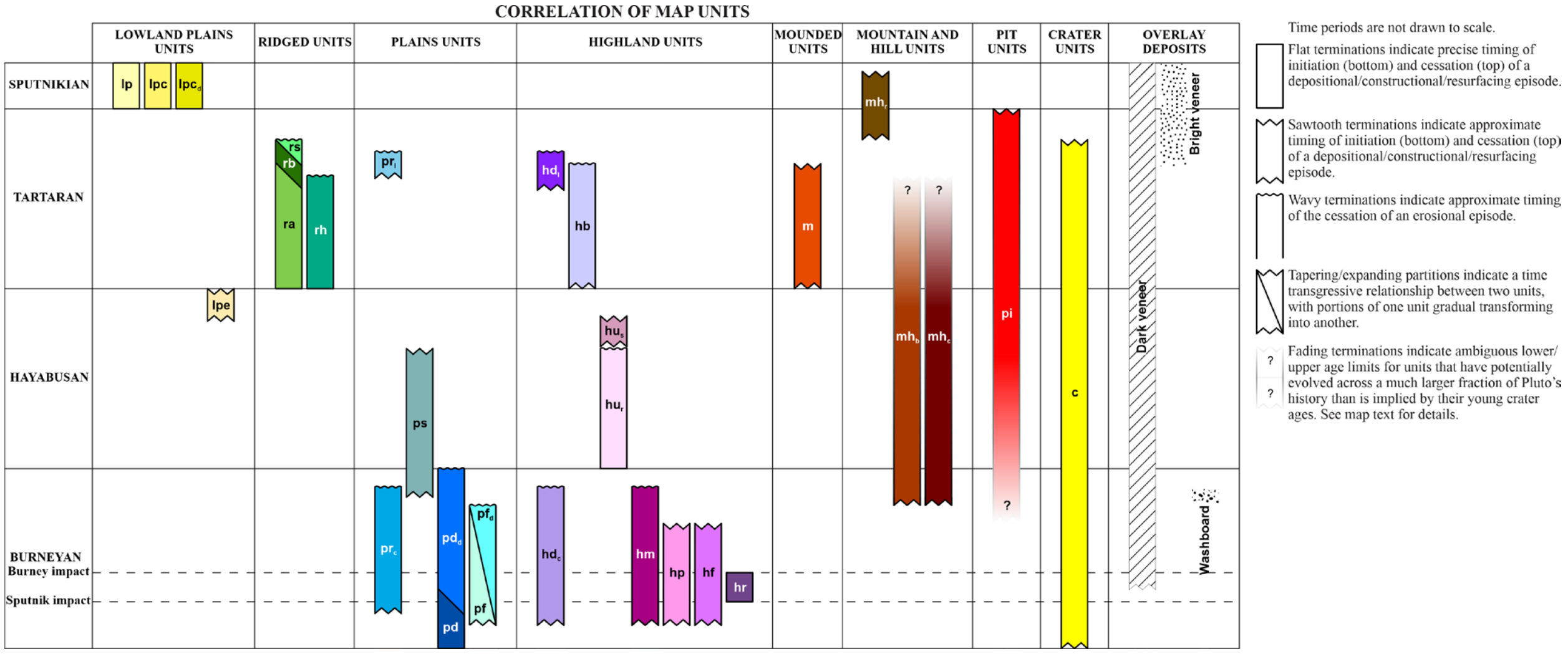

**Figure 2. Correlation of Map Units (COMU).** This figure from the Pluto geologic map is shown here for reference and displays all of the near-side units and their stratigraphic relationships.

The geologic map of Ganymede (Patterson et al., 2010; Collins et al., 2013) looked at cumulative crater densities for three diameter-based subsets of the mapped craters for some units or unit groups on the map. The three subsets were for craters >30 km in diameter, >20 km in diameter, and >10 km in diameter (the last was only presented in Patterson et al. [2010] and not on the map sheet). The display of several different crater diameter cutoffs attempted to mitigate the paucity of >30-km-diameter craters on some of the smaller-area units or unit groups, and also the fact that worse image resolutions in some areas may have affected a full count of smaller craters. Despite the data limitations, the trends were similar between the >30 km craters and the >20 km cumulative crater densities (Collins et al., 2013, their Figure 4). The relative crater densities were used as additional input along with superposition relationships to estimate the relative ages of units.

In this work we developed crater analysis techniques that are specific to the data available with USGS geologic maps. We apply these methods to estimate both relative and more quantitative ages for terrains using the current impact flux models for Pluto. We use the results to consider what percentage of Pluto's terrains have few-to-no craters and thus must have been resurfaced by one or more geologic mechanisms. We also discuss the specific mechanisms thought to be acting on each terrain (from both previous papers and our own interpretations for this work). The resurfacing mechanisms range from those that would be considered endogenic, such as volcanic, tectonic, or convective activity, to those considered exogenic processes like volatile sublimation or deposition.

## 2 Methods

### 2.1 Basemaps

The Long-Range Reconnaissance Orbiter Camera (LORRI), the Multispectral and Visible Imaging Camera (MVIC), and the Linear Etalon Imaging Spectral Array (LEISA) are the three main instruments that provided information useful for the geologic mapping and analysis described here (Cheng et al., 2008; Reuter et al., 2008; Howett et al., 2017; Weaver et al., 2020). The image basemap and topographic mosaic used for geologic mapping were built from both LORRI and MVIC panchromatic images (Schenk et al., 2018). We also used composition maps derived from the LEISA data for methane ($CH_4$), carbon monoxide (CO), nitrogen ($N_2$), water ice ($H_2O$), and a dark, red material generally understood to be a tholin mixture (Grundy et al., 2016; Protopapa et al., 2017; Schmitt et al., 2017; Protopapa et al., 2020; Cruikshank et al., 2021) to aid in unit definition and interpretation. Pluto has a rotation period of ~6.4 days and the higher resolution images were primarily taken within a 12-24 hour period near the spacecraft's closest approach. One hemisphere of Pluto was facing the spacecraft during the time of closest approach and was imaged at higher resolution (called the "encounter hemisphere" or "near-side" of Pluto, imaged at better than 1 km/pixel). The other sunlit areas of Pluto were imaged only at lower resolution during the approach phase when the spacecraft was farther away (called the "non-encounter hemisphere" or "far side" of Pluto).

The southern portion of Pluto, south of -30° latitude, was not directly illuminated at any point during Pluto's rotation period at this present point in its multi-million-year obliquity/precession cycles (Earle & Binzel, 2015; Young et al., 2021). However, some light was reflected by the hazes in Pluto's atmosphere into the night side of Pluto, thus revealing some additional terrain beyond the terminator (see **Figure 1**). These haze-lit areas extend down to approximately -55° latitude at the southernmost extent and have different illumination conditions and lower signal-to-noise. This image coverage was still very useful in extending our knowledge about Pluto's surface to a larger area, especially over the intriguing, likely cryovolcanic terrains.

## 2.2 Overview of Geologic Map Units

The overall geologic history of Pluto (for terrains currently observable) was broken into four time-stratigraphic periods, each named after a significant geologic feature associated with that era. The Burneyan, named after the Burney impact basin, encompasses the oldest terrains and is represented by mostly heavily cratered units with evidence of significant mantling and erosion. The Hayabusan, named after Hayabusa Terra, is an extensive middle-aged terrain interpreted to have formed through atmospheric deposition of fairly thick mantles, some of which are later eroded/altered with various scales and morphologies of pits. The Tartaran, named after Tartarus Dorsa, represents relatively young units (with few-to-no craters) with a variety of rough textures (appearing as ridge crests or peaks a few to 10 km apart) presumably produced through sublimation and/or other forms of erosion (Moore et al., 2018; White et al., 2021), as well as terrain south of Sputnik Planitia, which has been interpreted as likely cryovolcanic in origin. The Sputnikian includes the volatile ice units making up Sputnik Planitia and other, smaller volatile ice deposits in Pluto's equatorial regions, which are the very youngest terrains (no identifiable craters).

The geologic units were categorized into the following groups based on their morphologies and geographic settings (**Figures 1, 2, and S1**):

1. Lowland plains (lp, yellows) – volatile ice mantle materials, concentrated primarily within Sputnik Planitia, that have mostly experienced very recent resurfacing through sublimation into and deposition from Pluto's atmosphere, as well as glacial flow and convective overturn.
2. Ridged materials (r, greens) – primarily the Tartaran rough-textured units east of Sputnik Planitia, which are also thought to occur over large portions of the equatorial far side.
3. Plains (p, blues) – generally older, cratered, and low-relief regions that comprise most of the near side to the west of Sputnik Planitia.
4. Highlands (h, purples and pinks) – typically higher standing material, which can be plateau or mountain-like, that consists of predominantly older terrains, although some may have formed in Pluto's middle ages as they are less cratered.
5. Mounded materials (m, orangish red) – young, undulating terrain making up the likely cryovolcanic region to the southwest of Sputnik Planitia.
6. Mountains and hills (mh, browns) – crustal blocks ranging in size from a few km up to 50 km across, primarily located towards the edges of Sputnik Planitia.
7. Pits (p, red) – scalloped depressions often in quasi-linear arrangements concentrated east of Sputnik Planitia but also in the northern terrains and some to the southwest of Sputnik Planitia.
8. Craters (c, yellow) – impact craters.

9. Undifferentiated materials (u, blues) – intermediate and low albedo far side units where the image resolution is too low to distinguish between specific units.
10. Overlay deposits (hashing or stippling) – thinner deposits such as dark or bright veneers.

All far-side unit abbreviations are preceded by the letter “f”. The geologic map text and DOMU provides extensive descriptions of these units and we will refer to many of them below.

### 2.3 Crater identification for the geologic map

Two of the base map mosaics were the most useful for crater identification: the panchromatic mosaic and the topographic mosaic (both constructed from LORRI and MVIC images). Color and composition data (from the MVIC and LEISA instruments, respectively) were also consulted when appropriate. We identified craters for the map independently of previous work. Feature characteristics used to identify craters are:

- Negative topographic expression (a depression)
- Near-circular/elliptical planform outline
- Raised rim
- Proximal, radial ejecta deposits (not seen very often on Pluto because of erosion and also deposition of volatile ices; e.g., White et al., 2021; Singer et al. 2021)
- Interior structures associated with impacts, such as terraces, central peaks, or central pits

Many features across the surface of Pluto are considered highly likely to be impact craters, as they have many or all of the characteristics listed above and do not have characteristics of depressions produced through other possible mechanisms (e.g., erosion or collapse). In total 1329 craters in the 7-25 km diameter range were mapped. A small number of the features (~1-2%) have more ambiguous morphologies but still have several of the characteristics above, and were deemed more likely to be impact craters than other types of geologic features and thus were retained for this analysis. Lighting geometry, image quality (e.g., resolution and image signal-to-noise), and map projection can also play a role in crater identification on Pluto as described at length in previous work (e.g., Singer et al., 2019; Robbins & Singer, 2021; Singer et al., 2021; White et al., 2021) and we kept all of these in mind when mapping. For both the geologic mapping and crater identification, we also examined the images in their native geometry (not map projected), and/or used different map projections (e.g., Mercator vs polar), and tested different contrast stretches.

The lowest resolution images in the panchromatic base map mosaic used for mapping craters were 850 ± 30 m $px^{-1}$. The smallest features represented on the geologic map in its original draft were 7 km in diameter. Seven km also corresponds to ~8 pixels in the lowest resolution images (and would be a greater number of pixels across in any of the other, higher-resolution images in the base map), which is a reasonable number of pixels for crater identification given the geologically complex surface of Pluto (Singer et al., 2019; Singer et al., 2021). The smallest feature size on the geologic map was changed to 14 km during revisions to simplify the map. However, we chose to retain the 7 km lower limit for the crater analysis as the increased number of craters yielded more robust statistics. The craters in the 7 to 25 km size range were mapped only as point features on the geologic map, in accordance with the USGS mapping requirements for features below a certain scale, thus no specific diameter was available for these craters from this dataset. Craters larger than 25 km in diameter are not numerous on Pluto and there were often too few per unit for meaningful statistical analysis, thus they were not useful as a second bin.

2.4 Crater unit assignments

Craters in the dataset were then assigned to a mapped geologic unit (**Supplementary Table 1**). Unit assignments were initially made using the ArcMap function "Join Data", where the center point of each crater (from the geologic map point shapefile LocationFeatures) was used to pull the unit name from the spatially corresponding GeoUnits polygon shapefile (**Supplementary Data File 1**). Then every crater was checked to see if the assignment was appropriate and manually adjusted if not. The majority of the craters were fully contained within one unit and were assigned to that unit. When a crater fell on the boundary between two or more units, each case was considered individually. Typically, the crater was assigned to the presumed younger unit if that information could be inferred from superposition relationships, the erosional state of the two units, or other indications. The exception to this was if the younger unit could be seen to obviously embay or overprint the crater, in which case the crater was assigned to the underlying unit. We also carefully checked any craters that were assigned to units considered to be mantles, veneers, pitted terrain, or ponded volatile ice (i.e., a crater with volatile ice partially filling its floor), and adjusted their unit assignments where appropriate.

2.5 Crater data and plotting techniques

The area of each polygon in the GeoUnits shapefile was calculated (using geodesic areas from the "add geometry attributes" function in ArcMap) and the entire attribute table was exported. The attribute table for the crater points was also exported. The total area for each unit (the sum of all areas of the individual polygons of that unit; see **Table 1**) and the number of craters per unit were then used to calculate a measure of the density of craters per unit area. This is a common technique used to understand which terrains are older or younger. We used a crater density measure called R-values (the "R" stands for relative). We chose to use R-values because we wanted to use a binned (not cumulative) value, given that we had the number of craters in a specific range of diameters but no diameter information for each crater. Additionally, we chose to use R-values because corresponding age estimates were available for this format for Pluto and Charon (Greenstreet et al., 2015; 2016; and see the next section). Having just one diameter bin also simplified the age analysis to some degree (discussed below in **section 4.1**). Other crater value/plot formats (cumulative or differential) may be relevant for other types of crater data or other bodies.

To calculate R-values, the differential density of craters in a given size bin (not cumulative) is divided by the unit areas, and is also normalized to a distribution with $D^{-3}$, where −3 is a common power law slope found in crater size distributions and $D$ is the crater diameter. R-values ($R$) for each geologic unit are calculated as follows basaed on the standard technique (Crater Analysis Techniques Working Group, 1979):

$$R_{per_unit} = \frac{(N)}{(Bin_Width)*(Unit_Area)*D_{bin_center}{}^{-3}} \quad (1),$$

where $N$ is the number of craters per diameter bin in that geologic unit (in this case just one bin), *Unit_Area* is the total area of the given geologic unit as mapped across the surface of Pluto in km, *Bin_Width* is the bin upper diameter bound minus the bin lower diameter bound (in this case 25 − 7 = 18 km), and the $D_{bin_center}$ is the geometric mean of the bin (in this case 13.22 km). For the resulting plot, a differential distribution with a power law slope of −3 becomes a horizontal line, and slopes of −4 and −2 form 45° sloping lines (see dashed, blue lines in **Figure 3a**) on "square" plot where the x and y axes unit distances are equal distances physically on the plot. The one-sigma error bars are R-values divided by $\sqrt{N}$ (Crater Analysis Techniques Working Group, 1979).

Because we did not have diameter data for craters individually (as described above), we calculated the statistics for the one diameter bin available to us: from 7 to 25 km with a geometric mean of 13.22 km. Thus, all of our R-value points would fall on a single vertical line in a traditional R-plot (**Figure 3b**). Because our values were all for one bin with one mean diameter, the x-axis with a range of diameters for different bins was not needed, and we decided to make a new kind of visualization. The new visualization, which we call a distributed R-plot, takes each R-value for each unit from a vertical "cut" at 13.22 km diameter from the traditional R-plot, spreads them out horizontally, and orders them in the same sequence as the COMU (**Figure 3c, Figure 4**). The distributed R-plot simplifies the number of "dimensions" from three in the traditional R-plot in **Figures 3a,b** (crater diameter, R-value, and age predictions) to only two in **Figure 3c** (R-value and age predictions). This distributed R-plot allows individual points to be more easily seen and their relative R-values more easily compared to the predicted age lines (see description below). Additionally, the distributed R-plot allows the points to be more easily grouped by unit type. The color for each point on the distributed R-plot was matched to the unit color in the geologic map. We chose to plot all units with 2 or more craters, but indicated units with less than 10 craters by giving their points gray outlines (instead of black).

**Table 1: Near-side units**

| Near-side Unit Group | Unit abbr. (in DOMU order) | Unit name | Unit color on map | Total unit area (km^2) | Number of craters between 7 and 25 km* | Number of craters > 25 km* | Notes on craters |
|---|---|---|---|---|---|---|---|
| LOWLAND PLAINS UNITS | lp | Lowland plains material | | 3.967E+05 | 0 | 0 | No obviously superposing craters |
| | lpc | Cellular lowland plains material | | 4.475E+05 | 0 | 0 | No obviously superposing craters |
| | $lpc_d$ | Dark, cellular lowland plains material | | 4.089E+04 | 0 | 0 | No obviously superposing craters |
| | lpe | Etched, cratered lowland plains material | | 3.394E+04 | 2 | 0 | Possibly one superposing large crater (> 25 km) along an edge of this terrain, but difficult to tell the superposition relationship |
| RIDGED UNITS | ra | Arcuate ridged material | | 1.895E+05 | 0 | 0 | No craters appear to be superposing this unit, although a few may be superposed by the unit itself |
| | rb | Bladed ridged material | | 4.042E+05 | 0 | 0 | No obviously superposing craters |
| | rs | Subdued ridged material | | 1.742E+05 | 1 | 0 | One crater between 7-to-25 km near edge of terrain, no obviously superposing large craters > 25 km |
| | rh | Hummocky ridged material | | 2.011E+05 | 5 | 0 | One large crater (> 25 km) interior to this unit, but it appears to be eroded similar to the rest of the terrain |
| PLAIN S UNITS | $pr_l$ | Lightly cratered rough plains material | | 7.849E+04 | 0 | 0 | No obviously superposing craters |
| | $pr_c$ | Cratered rough plains material | | 2.446E+05 | 95 | a few | Up to 5 large craters (> 25 km), but some are on the edges and difficult to tell if they overprint |
| | ps | Smooth plains material | | 2.260E+05 | 26 | a few | A few possibly superposing large craters (> 25 km), but mostly this unit appears to instead be overprinting most large craters |
| | pd | Dark plains material | | 9.423E+05 | 244 | many | Many large craters (> 25 km) superpose this unit |
| | $pd_d$ | Denuded dark plains material | | 8.640E+04 | 13 | 0 | One large crater (> 25 km) near the edge of the terrain but it is not obviously superposing |
| | pf | Fretted plains material | | 1.363E+06 | 396 | many | Many large craters (> 25 km) superpose this unit |
| | $pf_d$ | Denuded fretted plains material | | 3.570E+05 | 95 | moderate amount | Up to ~12 likely superposing large craters (> 25 km), again somewhat difficult to discern the superposition relationships |
| **HIGHLAND UNITS** | $hd_l$ | Lightly cratered dissected highland material | | 4.003E+04 | 1 | 0 | One possible large superposing crater (> 25 km) along an edge of this terrain, but difficult to tell the superposition relationship |
| | $hd_c$ | Cratered dissected highland material | | 1.210E+05 | 19 | a few | Four large craters (>25 km) may superpose this unit, but they also may be partially modified by this terrain |
| | hb | Blocky highland material | | 1.359E+05 | 0 | 0 | No obviously superposing craters |
| | $hu_s$ | Smooth, undulating highland material | | 7.105E+04 | 9 | a few | A few (2-3) likely superposing large craters (>25 km), again somewhat difficult to discern the superposition relationships |
| | $hu_r$ | Rough, undulating highland material | | 6.748E+05 | 74 | moderate amount | Up to ~10 likely superposing large craters (> 25 km), again somewhat difficult to discern the superposition relationships |
| | hm | Dissected massif highland material | | 3.748E+05 | 76 | moderate amount | Up to ~6 likely superposing large craters (> 25 km), again somewhat difficult to discern the superposition relationships |
| | hp | Polar highland material | | 3.511E+05 | 62 | moderate amount | Up to ~10 likely superposing large craters (> 25 km), again somewhat difficult to discern the superposition relationships |
| | hf | Rugged, fractured highland material | | 2.321E+05 | 78 | moderate amount | Up to ~12 likely superposing large craters (> 25 km), again somewhat difficult to discern the superposition relationships |
| | hr | Cratered rough highland material | | 2.427E+05 | 132 | a few | 3-4 likely superposing large craters (> 25 km) |
| MOUD ED UNITS | m | Hummocky mounded material | | 2.003E+05 | 1 | 0 | The one small crater is near the edge of this unit |
| MOUNTAIN AND HILL UNITS | $mh_r$ | Rubbly hill material | | 4.336E+03 | 0 | 0 | No obviously superposing craters |
| | $mh_b$ | Blocky mountain material | | 1.178E+05 | 0 | 0 | No obviously superposing craters |
| | $mh_c$ | Corrugated hill material | | 1.293E+04 | 0 | 0 | No obviously superposing craters |
| PIT UNITS | pi | Pit material | | 1.949E+05 | N/A | N/A | |
| CRAT ER UNITS | c | Crater material | | 2.852E+05 | N/A | N/A | |
| **Totals:** | | | | **8.245E+06** | **30** | | |
| **Percentage of Pluto's Surface Area:** | | | | **46%** | | | |

*Cells with a gray background are those with zero or one craters and are shown in **Figure 5**.

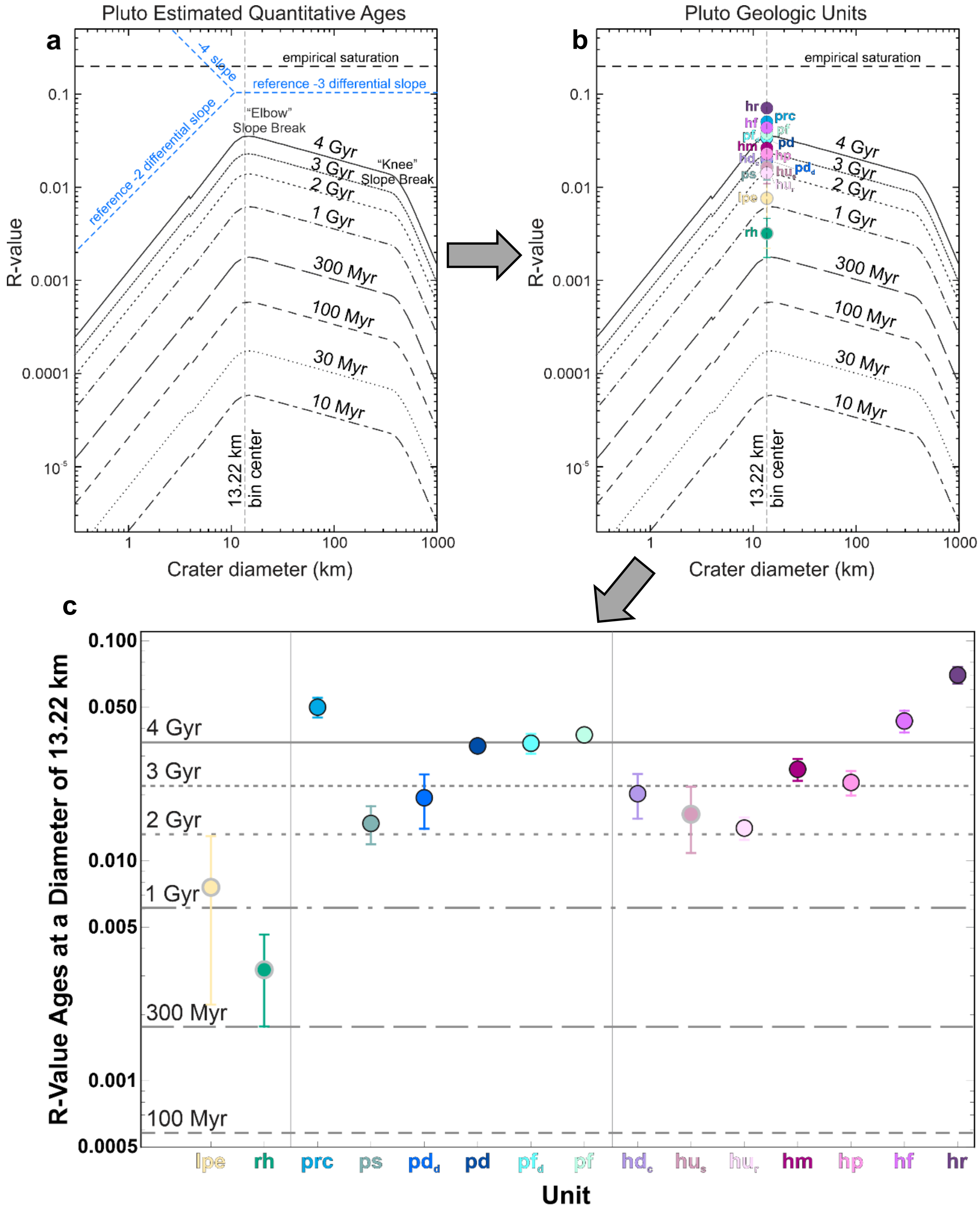

**Figure 3. Concept of distributed R-plots.** (a) Basic R-plot with reference estimated age lines (Greenstreet et al., 2015; Singer et al., 2021) for Pluto (black curves) and reference differential slopes (blue, dashed lines) overlain. (b) Pluto geologic units with two or more craters between 7 and 25 km in diameter shown in a traditional R-plot format, also with age lines overlain. (c) Pluto geologic units shown in a new distributed R-plot format. See text for details.

## 3 Results: Unit crater densities and relative ages

Out of the 36 total units on the geologic map (including the 6 far side units), 12 units contained more than 10 craters, 3 units contained between 2 and 10 craters, and 21 units had one or zero craters. Relative ages can be fairly easily compared between the 15 units with 2 or more craters as shown on the distributed R-plot. Relative ages from comparing average crater densities per unit are a valuable contribution to understanding the sequence of geologic events on a boy, and do not require an extra step of converting to a quantitative age.

### 3.1 Lower crater density/intermediate-age or younger units

The unit with the lowest crater density on the distributed R-plot is the hummocky ridged material (rh, teal), which is a topographically rough terrain along the southeastern border of the encounter hemisphere imaging (see **Figure 1**). It is a relatively large unit spanning the dark Safronov Regio and extending into the hazelit region of the basemap (and likely extends onto the far side continuing in Sharaf, Harrington, and Belton regios). There are five craters visible in the hazelit region of this unit. The large area and relatively small number of recognizable craters give this unit its low density relative to the other cratered terrains. While unit rh is clearly eroded and at least partially resurfaced, the existence of some larger craters (>20 km in diameter) indicates this is one of the few intermediate-aged terrains (as opposed to the very young regions with no craters and the older regions with many craters). The craters on this unit allowed us to infer that the other ridged units (rX, greens) are likely somewhat older than the nearby mounded units (m), as there were otherwise no clear superposition relations at the boundary between the two unit groups, both of which define the Tartaran period in Plutonian history.

The second lowest R-value is for the etched, cratered lowland plains material (lpe, light yellow). This unit occurs in broad low-elevation patches, primarily in the bottom of large depressions >30 km across, some of which may be large eroded craters (e.g., Bower or Kowal craters) in the northeastern mid-latitudes of the encounter hemisphere and is the only lowlands plains unit with superposing small craters (<10-km-diameter). This unit has previously been noted as unique because it has some features associated with volatile ice composition ($N_2$) but unlike Sputnik Planitia also retains some intermediate-scale topography (White et al., 2021; White et al., 2023). Thus, again this is likely an intermediate-age terrain on Pluto.

### 3.2 Higher crater density/older units

The highest crater density unit is the cratered rough highlands material (hr, dark purple). The hr unit spans some regions that appear very eroded (including some rugged, shallow, 30-50-km-diameter craters), and also includes the ~250-km-diameter Burney basin. The Burney basin has previously been determined to be an ancient feature with one of the highest crater densities on Pluto's encounter hemisphere (Singer et al., 2021; White et al., 2021). Thus, this is confirmation that the range of crater diameters used here produces a similar relatively high crater density to those derived from looking at a wider range of crater sizes in previous work.

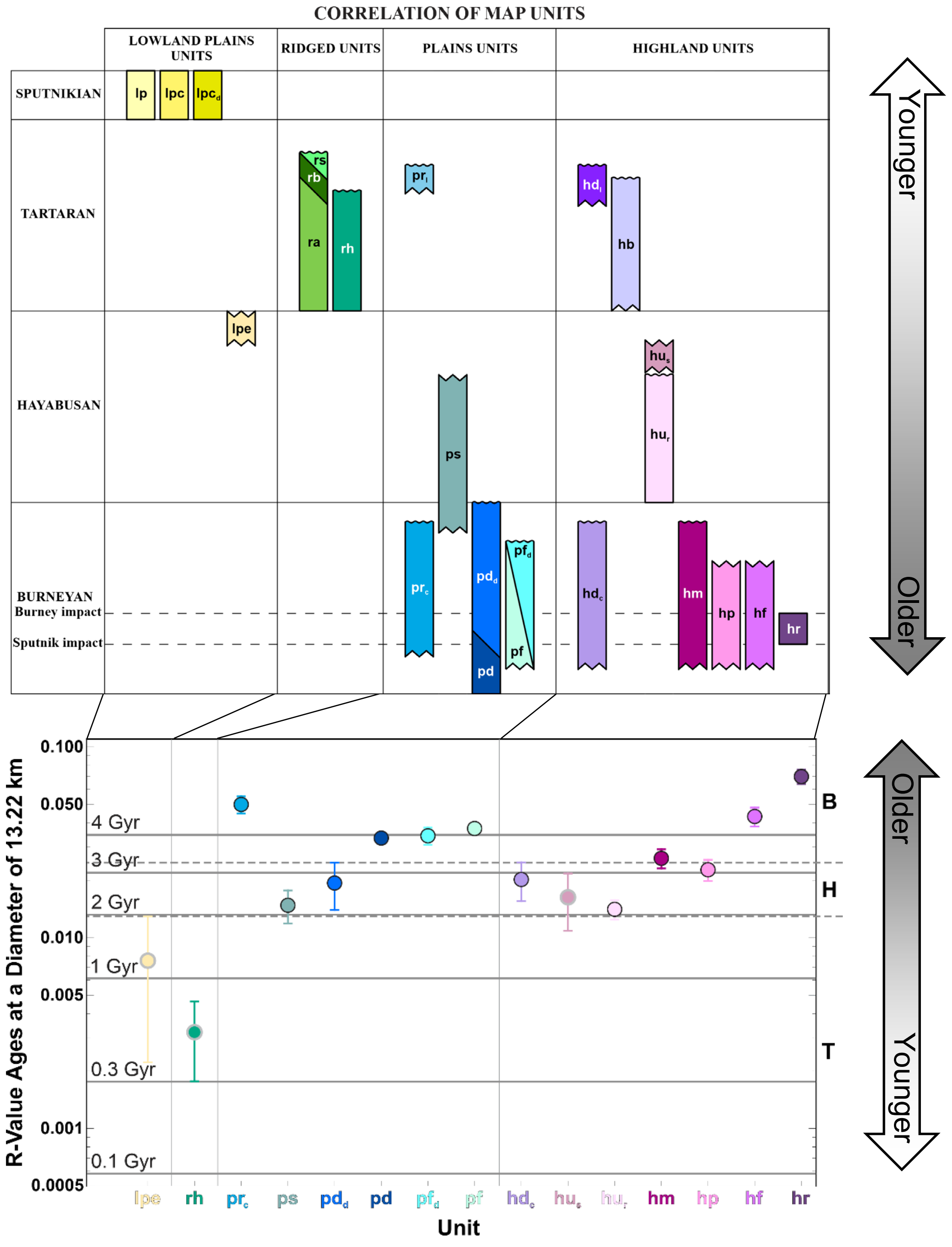

**Figure 4. Distributed R-plot and relative age on the COMU.** The crater spatial densities, as displayed on the R-plot as R-values, provide a measure of the relative ages of the intermediate-aged and older terrains on Pluto. Many units have similar values within the error bars, but some terrains are much more lightly or heavily cratered. The relative crater densities for the units generally agreed with other indicators of relative age, such as superposition. The estimated quantitative ages (solid lines) in the distributed R-plot are provided for reference, although they are estimated to have a factor of 2 uncertainty. The geologic periods are also indicated on both plots. See text for details.

All of the plains units (pX, blue and green-toned) and highlands units (hX pink and purple-toned) have relatively high R-values. Quite a few of the R-values for these units are similar to each other considering the error bars. But there is still a factor of ~five or more difference between the lowest and highest R-values for these unit groups, which could still account for a few billion years according to the current impactor flux estimates. In general, the crater densities for the plains and highlands units match what would be expected for the relative ages based on unit superposition and apparent crater degradation levels (White et al., 2021; Hedgepeth et al., 2023).

## 4 Discussion

### 4.1 Quantitative age estimates

Quantitative age estimates (e.g., 100 Ma or 2 Ga) for surface units can also be useful as comparisons to and constraints on dynamic and interior evolution modeling. Here we describe our general philosophy and caveats to estimating quantitative ages, and the method for deriving the age lines in **Figure 3**.

Age estimates for a given geologic unit based on crater statistics are always average ages, as even a unit that appears to be fairly homogeneous could have been formed at different times or experienced spatial heterogeneous erosion/resurfacing. The assignment of ages follows the same general philosophy as the identification of units, where an attempt is made to identify regions of a planetary surface that formed by the same general process and are composed of the same materials. In some cases, it is difficult to tell how thick a unit is, but as mentioned above we paid special attention to any unit that might be considered a mantle or veneer. Typically craters superpose the unit they were assigned to, and there were only a few ambiguous cases where the unit appeared to superpose the crater instead. Additionally, the impact cratering process itself resurfaces, but the ages again typically refer to the unit the crater formed in. One exception on the Pluto map is for the Burney basin, which makes up a large fraction of the cratered rough highlands material (hr, dark purple, described above as the unit with the highest R-value). The basin covers a large area and is itself superposed by many smaller craters, making it possible to date the basin itself (Singer et al., 2021), or the basin as part of the hr terrain, as was done here.

In addition to comparing the crater spatial densities between units to understand which are relatively older or younger units, we also utilized the expected crater densities for a given age surface on Pluto as calculated by Greenstreet et al. (2015, 2016). The expected crater densities for a given age surface are derived from the impact probabilities of different Kuiper belt object (KBO) subpopulations onto Pluto and Charon over the past ~4 billion years. The overall number of impactors per size (per year) was based on the population of known KBOs at larger sizes observed by telescopic surveys (primarily KBOs larger than ~50-100 km in diameter). The estimates were then extrapolated to smaller objects, as most of the craters on Pluto are made of KBOs/impactors <40 km in diameter. After the New Horizons flyby, this extrapolation was adjusted so that the size-distribution slope for the smaller impactors/craters matched the size-distribution slope of smaller craters as observed on Pluto and Charon (Singer et al., 2019; Singer et al., 2021), but the overall magnitude of the impact flux was the same as in earlier works. The age lines shown in **Figure 3a,b** change slope based on the crater size-frequency distribution from the New Horizons data but remain parallel. By coincidence, the middle of our one diameter bin, 13.22 km, falls near the transition in slopes between larger craters with a steeper SFD slope, and smaller craters with a shallower SFD slope (this transition was nicknamed the "elbow" break in slope; see **Figure 3**). The estimated R-values for 8 specific ages at our middle bin diameter are shown in **Figure 3** and **Table 2**.

**Table 2. Age lines R-values**

| Age | Estimated R-value at a diameter of ~13.22* |
|---|---|
| 4 Gyr | 0.034581 |
| 3 Gyr | 0.021924 |
| 2 Gyr | 0.013213 |
| 1 Gyr | 0.006125 |
| 300 Myr | 0.001759 |
| 100 Myr | 0.000580 |
| 30 Myr | 0.000173 |
| 10 Myr | 0.000058 |

*Note we include 5 significant figures only to be able to easily compare these numbers to **Figure 3c**, not because there is that much precision in the estimate. Please see section 2.5 for a description of the uncertainties in these estimates.

There are a number of sources of uncertainty in the age estimates due to incomplete knowledge of the Kuiper belt and its subpopulations. The age estimates do account for some collisional decay of the impacting population over the past 4 billion years but do not estimate any early phase (prior to 4 billion years ago) of higher bombardment that may have occurred before or during Pluto's emplacement in its current orbit (Greenstreet et al., 2015). In previous works, the age uncertainty was estimated to be a factor of two (Greenstreet et al., 2015; Singer et al., 2021). If some percentage of the impacting population is made up of binary objects, that could also shift the predicted ages (Parker, 2021; Singer et al., 2021) to lower values. If some of the impactors in the calculated impact rate result in two craters, that means less time is needed to accumulate the same number of craters. Secondary craters are not evident on Pluto (no obviously clustered sets of craters or rays). Additionally, secondary craters would not be theoretically abundant at crater sizes greater than 7 km. The largest secondaries are typically 5-8% of the primary size. Thus, primary craters greater than ~100 km would be needed to produce substantial numbers of secondaries above our 7 km diameter size cutoff. There are only four near side craters/basins larger than ~100 km and none of them appear to be fresh or recent (the four are the craters/basins named Oort, Edgeworth, Burney, and Sputnik).

After the completion of the Pluto geologic map, a revised impactor flux was calculated by Greenstreet et al. (2023). The expected levels of impactor flux increased by ~a factor of two primarily because of an increase in the estimate of the total number of hot classical KBOs. With an increased impact flux, smaller amounts of time are needed to reach a given crater density, and thus the estimated age for a given crater density decreases also by a factor of ~2.

To extrapolate between the age lines shown in **Figure 3**, we fit a simple polynomial (**Figure S2**) to the values in **Table 2**. The estimated ages as a function of R-value ($R$) for a crater diameter of 13.22 km is: $-1.6716\text{E}\text{x}10^{12}\times R^2 + 1.7348\text{x}10^{11}\times R + 1.6426\ \text{x}10^4$. This function was used to estimate the values in **Table 3**.

We used a conservative approach in the geology map text and previous work (e.g., Singer et al., 2021), where we used the approximate value at the top of the error bar (the highest/oldest value) to give an effective maximum age estimate (**Table 3**). For younger units with few craters, and thus large error bars, these units could be substantially younger than this maximum age, so again these should not be taken as a precise "age". Additionally, even this upper limit age could have approximately a factor of 2 uncertainty in it. Given that the new values in Greenstreet et al. (2023) would revise these maximum ages downward (Greenstreet et al., 2023), we have left the upper limit as-is without increasing it even further.

Given the large error bars on some of the R-values, in combination with the uncertainty factors on the age estimates, we did not feel it was appropriate to pull out specific ages for each of the point R-values for the terrains (i.e., state exactly what ages the R-value points fall on in **Figure 3**) as this implies a higher level of precision than we can currently calculate. As the knowledge of Kuiper belt populations increases, more specific ages may be possible.

**Table 3. Unit R-values and ages**

| Unit | Number of craters | Total unit area (km^2) | R-value | R-value error bar (1σ) | Upper limit R-value* | Upper limit age (yrs)* | Upper limit age in Gyrs* |
|---|---|---|---|---|---|---|---|
| $lp_e$ | 2 | $3.394\text{x}10^4$ | 0.00758 | 0.00536 | 0.01294 | $< 1.96\text{x}10^9$ | < 2.0 |
| rh | 5 | $2.011\text{x}10^5$ | 0.00320 | 0.00143 | 0.00463 | $< 7.67\text{x}10^8$ | < 0.8 |
| $pr_c$ | 95 | $2.446\text{x}10^5$ | 0.04995 | 0.00513 | 0.05508 | $< 4.48\text{x}10^9$ | < 4.5 |
| ps | 26 | $2.260\text{x}10^5$ | 0.01480 | 0.00290 | 0.01770 | $< 2.55\text{x}10^9$ | < 2.5 |
| $pd_d$ | 13 | $8.640\text{x}10^4$ | 0.01935 | 0.00537 | 0.02472 | $< 3.27\text{x}10^9$ | < 3.3 |
| pd | 244 | $9.423\text{x}10^5$ | 0.03330 | 0.00213 | 0.03544 | $< 4.05\text{x}10^9$ | < 4.0 |
| $pf_d$ | 95 | $3.570\text{x}10^5$ | 0.03423 | 0.00351 | 0.03774 | $< 4.17\text{x}10^9$ | < 4.2 |
| pf | 396 | $1.363\text{x}10^6$ | 0.03738 | 0.00188 | 0.03925 | $< 4.23\text{x}10^9$ | < 4.2 |
| $hd_c$ | 19 | $1.210x10^5$ | 0.02020 | 0.00463 | 0.02483 | $< 3.28x10^9$ | < 3.3 |
| $hu_s$ | 9 | $7.105x10^4$ | 0.01629 | 0.00543 | 0.02172 | $< 2.98x10^9$ | < 3.0 |
| $hu_r$ | 74 | $6.755x10^5$ | 0.01409 | 0.00164 | 0.01573 | $< 2.32x10^9$ | < 2.3 |
| hm | 76 | $3.748x10^5$ | 0.02608 | 0.00299 | 0.02907 | $< 3.63x10^9$ | < 3.6 |
| hp | 62 | $3.511x10^5$ | 0.02271 | 0.00288 | 0.02559 | $< 3.35x10^9$ | < 3.3 |
| hf | 78 | $2.321x10^5$ | 0.04322 | 0.00489 | 0.04812 | $< 4.48x10^9$ | < 4.5 |
| hr | 132 | $2.427x10^5$ | 0.06994 | 0.00609 | 0.07603 | $< 3.53x10^9$ | < 3.5 |

*Value for the upper error bar location (the point estimate + 1σ) and for age estimates given by Greenstreet et al. (2015) and Singer et al. (2021). Many terrains could be *substantially younger* than this upper limit. See text for additional details on R-values and age estimate uncertainties.

### 4.2 Younger units with one or zero craters

As the impactor flux models become more refined in the future, methods like those described in Michael et al. (2016) to estimate ages for terrains with small numbers of craters could be employed for the Pluto units with one or zero craters. However, we can still look at the simple metrics to derive some information about these units. Some regions on Pluto with few-to-no craters were given upper limit ages in Singer et al. (2021). These upper limit age estimates were done before the geologic map, and so do not directly correspond to the units on the map.

For units mapped on Pluto that show no craters larger than 7 km in diameter, the surface area of the unit can serve as a proxy for relative age. The larger the area, the more improbable it is that no craters would form in a given length of time (e.g., Singer et al., 2021). Thus, larger units with no craters are likely to be younger than (or at least the same age as) smaller ones with no craters.

For the 13 near-side units with one or zero craters (excluding the pit unit and the crater unit), 10 have zero craters, and 3 have one crater. For the terrains with one crater, the crater is near the edge of the unit and it is not always clear if it is completely superposing the unit. Thus, for the simple analysis conducted here, we treat all 13 units with one or zero craters the same way. These 13 units make up about 27% of the surface area of the mapped encounter hemisphere units (or about ~13% of the total surface area of Pluto).

**Figure 5** displays the areas for each of the 13 units with one or zero craters. As always when dealing with "small number" statistics, this area comparison does not provide strong constraints but is likely more meaningful for the larger-area units. Many of these 13 units have small areas (less than about 200,000 $km^2$ or less than ~2.5% of the encounter hemisphere total area), and the units are sometimes a "subset" of a larger unit based on albedo (e.g., the $lpc_d$ dark plains unit). Thus, in general, the units with small areas are difficult to conclude much about.

Therefore, the main insight we take from **Figure 5** refers to the units with larger areas: the bladed ridged material (rb) may be a very young terrain, as its area is similar to that of some of the Sputnik Planitia plains units (lp or $lp_c$). Although the surface area of the near-side bladed ridged material (rb) is less than that of the Sputnik Planitia units combined ($lp+lp_c$), the bladed unit very likely continues for vast expanses on Pluto's far-side.

The bladed texture of this terrain and the general lack of obvious craters already indicated substantial resurfacing (Moore et al., 2018; White et al., 2021; Young et al., 2021). The analysis presented here further suggests that this unit has undergone extensive resurfacing in Pluto's relatively recent geologic past to maintain a crater-free area, which has presumably involved large amounts of sublimation and deposition of volatile ices (Moore(White et al., 2021; Young et al., 2021) et al., 2018). Singer et al. (2021) found an upper limit age for the entire area of Sputnik Planitia of less than ~50 Million years. Thus, Pluto displays more than one type of terrain that may be as young as some of the very active outer Solar System moons like Europa (Bierhaus et al., 2009), Triton (Stern & McKinnon, 2000; Schenk & Zahnle, 2007; Mah & Brasser, 2019; McKinnon et al., 2024), and Enceladus (e.g., Porco et al., 2006; Spencer & Nimmo, 2013; Nimmo et al., 2018; Patterson et al., 2018).

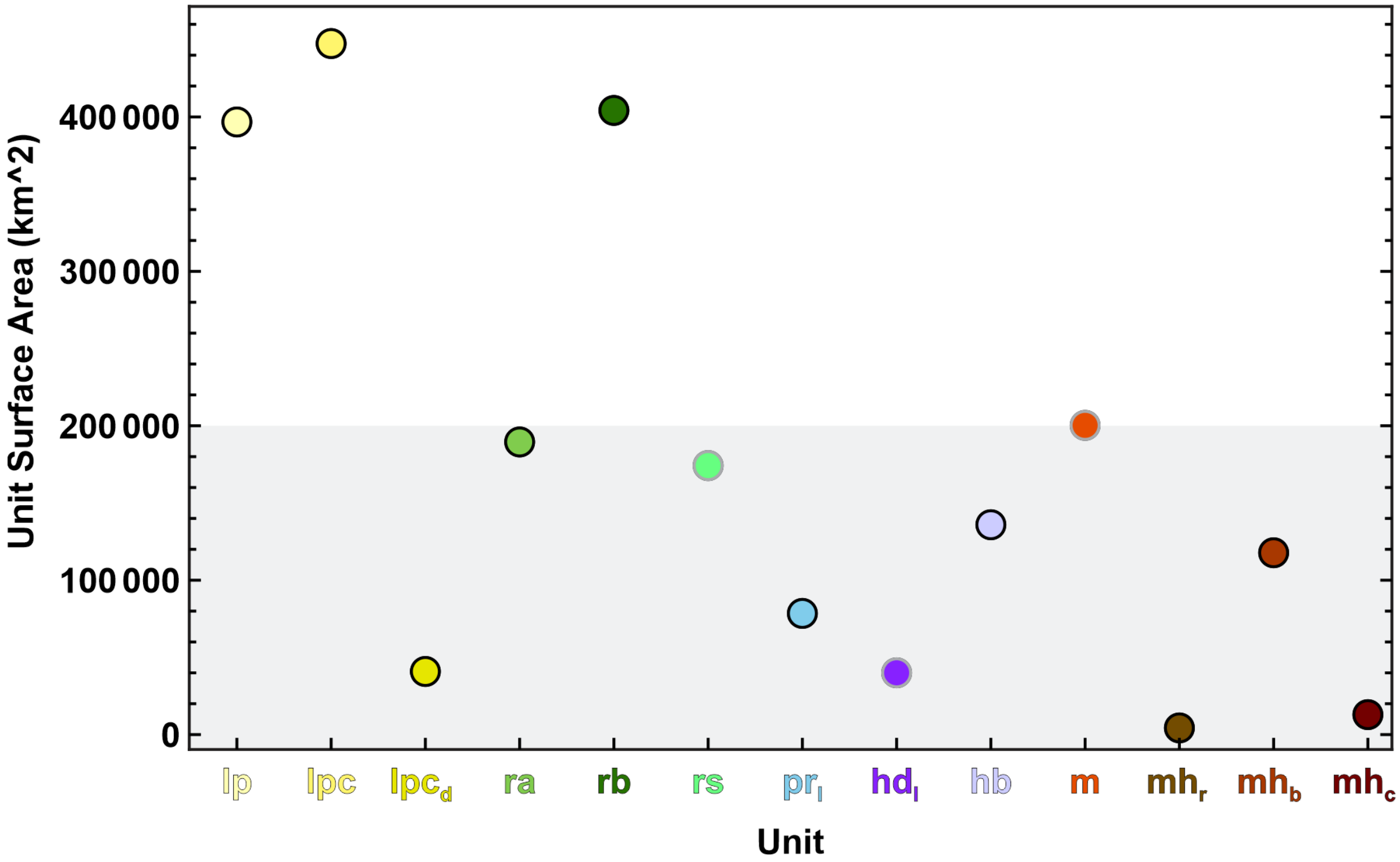

**Figure 5. Unit surface areas for units with one or zero craters.** The three large area units (lp, lpc, and rb) are all potentially very young, as the larger the area the more unlikely it is a crater would not have formed in Pluto's recent past. Units with zero craters are shown as a colored point with a black outline, the three units with one crater near the edge of the unit are shown with a grey outline. The lower part of the plot is greyed to indicate these are all very small units (< ~2.5% of the mapped encounter hemisphere surface area).

## 5 Conclusions

We developed several methods for analyzing crater data specific to USGS geologic maps, in which smaller craters are mapped only as point features. Even with just the number of craters per unit (in the small crater size bin of 7-to-25 km) we were able to supplement other indicators of terrain relative age such as superposition and terrain degradation. We also laid a foundation for estimating quantitative ages (e.g., 100 Ma or 2 Ga) on Pluto. Future models of Kuiper belt subpopulations and their impact rates onto Pluto can be applied to update the quantitative age estimates. Each Solar System body has a unique set of image data and geology, but the techniques developed here may be applicable to future USGS geologic map crater data.

Pluto's geologic terrains span a wide range of ages, but also display diverse surface textures within an age group. There are several types of older, heavily-cratered terrains, including both plains and highlands units (lpe and rh) covering ~64% of the surface area of the mapped near-side. Terrains that could be considered intermediate-aged on Pluto are somewhat less frequent, but also quite different from each other; the smooth plains and hummocky ridged material make up ~3% of the near-side. Intriguingly young terrains, resurfaced through both endogenic and exogenic processes, are found with a wide variety of morphologies. Young units (including all those shown in **Figure 5** and the pitted terrain) make up ~30% of the near-side and potentially a large fraction of the far-side if the bladed terrains do indeed extend there (Moore et al., 2018; Stern et al., 2021). The larger craters (> 25 km in diameter) mapped as the crater unit make up the remaining ~3%.

## Acknowledgments

This work was done in support of the creation of a global geologic map of Pluto, which was funded by the NASA Planetary Data Archiving, Restoration, and Tools (PDART) grant 80NSSC19K0423. RMCL's work was done at the Jet Propulsion Laboratory, California Institute of Technlogy, under contract with NASA. We also thank two reviewers for their comments that helped improve this manuscript.

## Open Research

All measurements and crater data displayed above are available as either in-text tables or in **Supplementary Table S1**. Table S1 is also available as a figshare.com respository at https://doi.org/10.6084/m9.figshare.27640317.v1 under at CC BY 4.0 license (Singer et al., 2024). Table S1 includes the list of small crater locations (latitudes and longitudes) and assigned geologic units. The basemap mosaics, topography, and composition maps that correlate with the crater locations provided are available from the Planetary Data Systems Small Bodies Node https://pds-smallbodies.astro.umd.edu/data_sb/missions/newhorizons/index.shtml in the "New Horizons Pluto Encounter Geology and Geophysical Maps" and "New Horizons Pluto Encounter Surface Composition Maps" directories.

Supporting Information for

**Pluto Geologic Map: Use of Crater Data to Understand Age Relationships**

K. N. Singer[1], O. L. White[2], S. Greenstreet[3], J. M. Moore[4], D. A. Williams[5], and R. M. C. Lopes[6]

[1]Southwest Research Institute, Boulder, Colorado, USA. [2] Carl Sagan Center at the SETI Institute, Mountain View, California, USA. [3]Department of Astronomy and the DiRAC Institute, University of Washington, Seattle, Washington, USA. [4] NASA Ames Research Center, Moffett Field, California, USA. [5] Schoolof Earth & Space Exploration, Arizona State University, Tempe, Arizona, USA. [6] Jet Propulsion Laboratory, California Institute of Technology, Pasadena, CA, USA.

## Contents of this file

Figures S1 to S2

## Additional Supporting Information (Files uploaded separately)

Caption for Table S1

(a) Lowland plains units

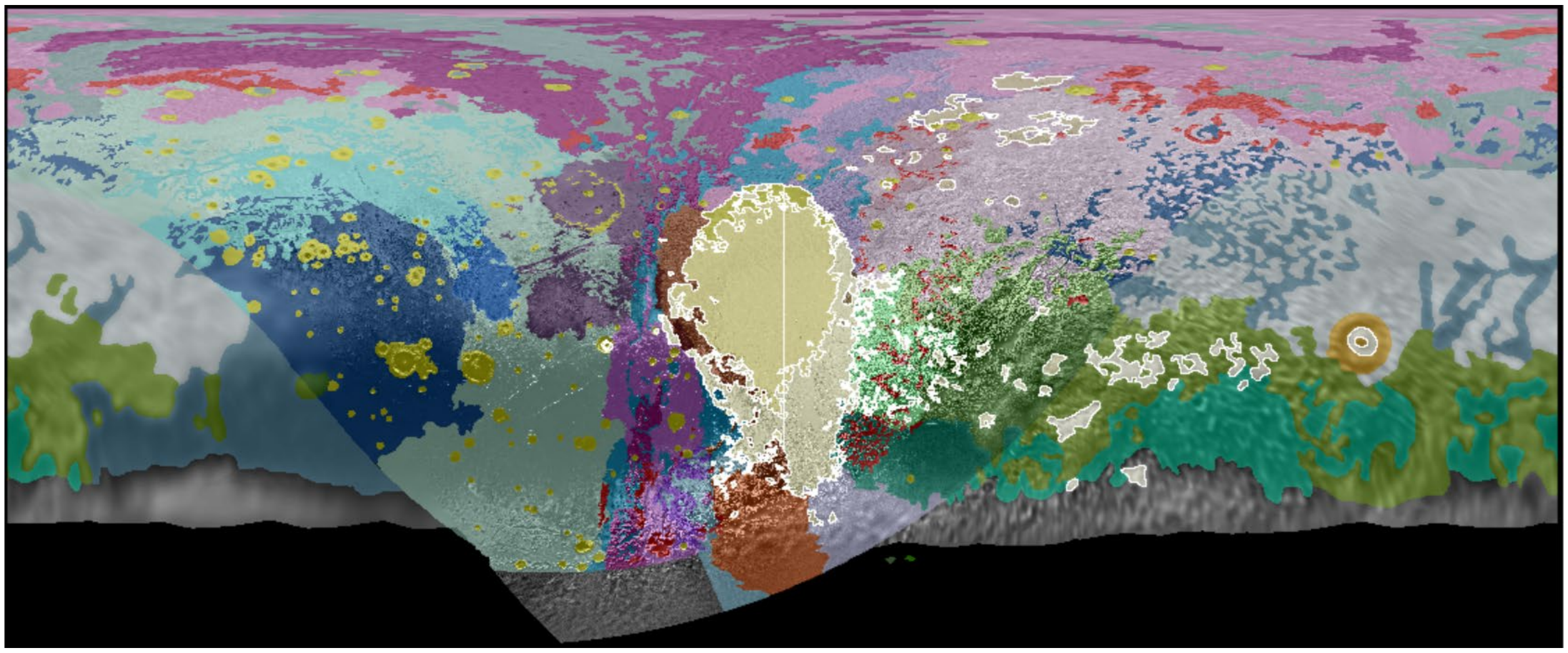

(b) Ridged units

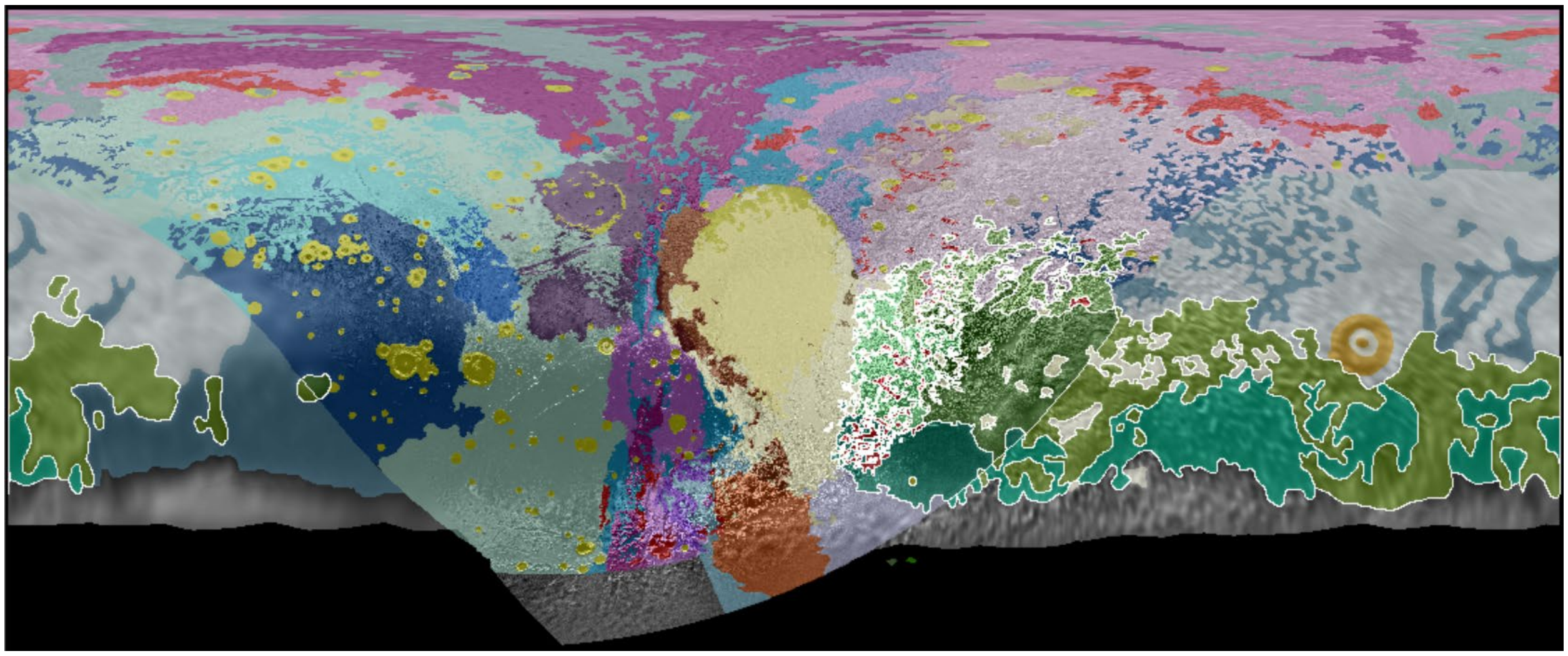

(c) Plains units

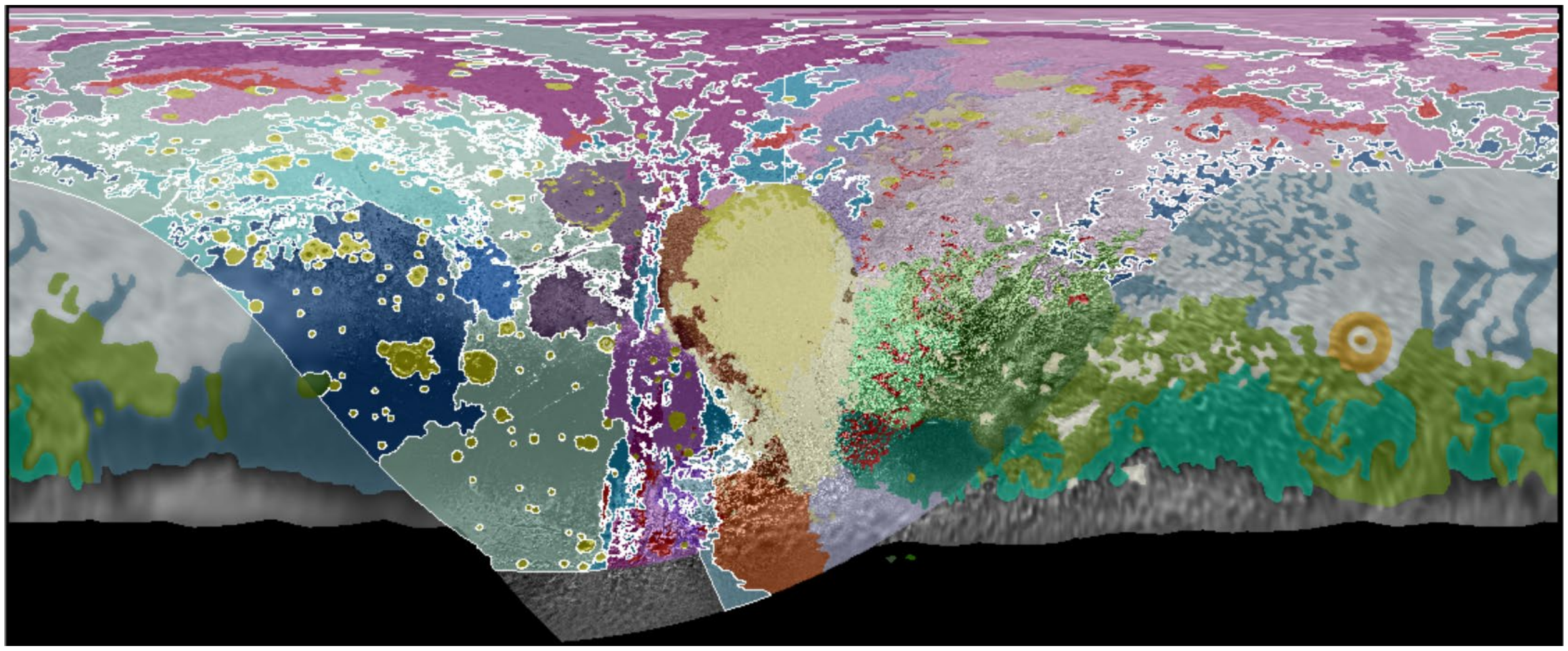

(d) Highlands units

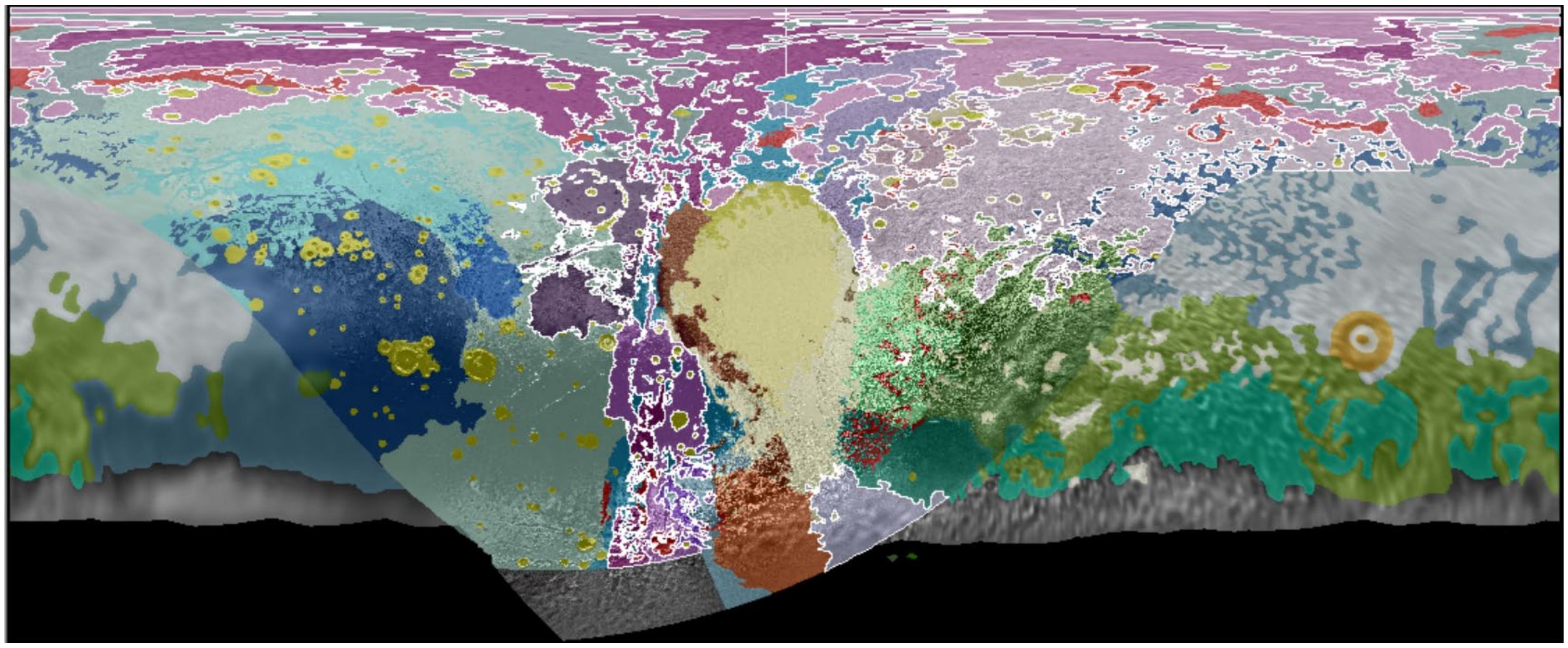

**Figure S1 a-d. Highlighted unit groups.** See text for details.

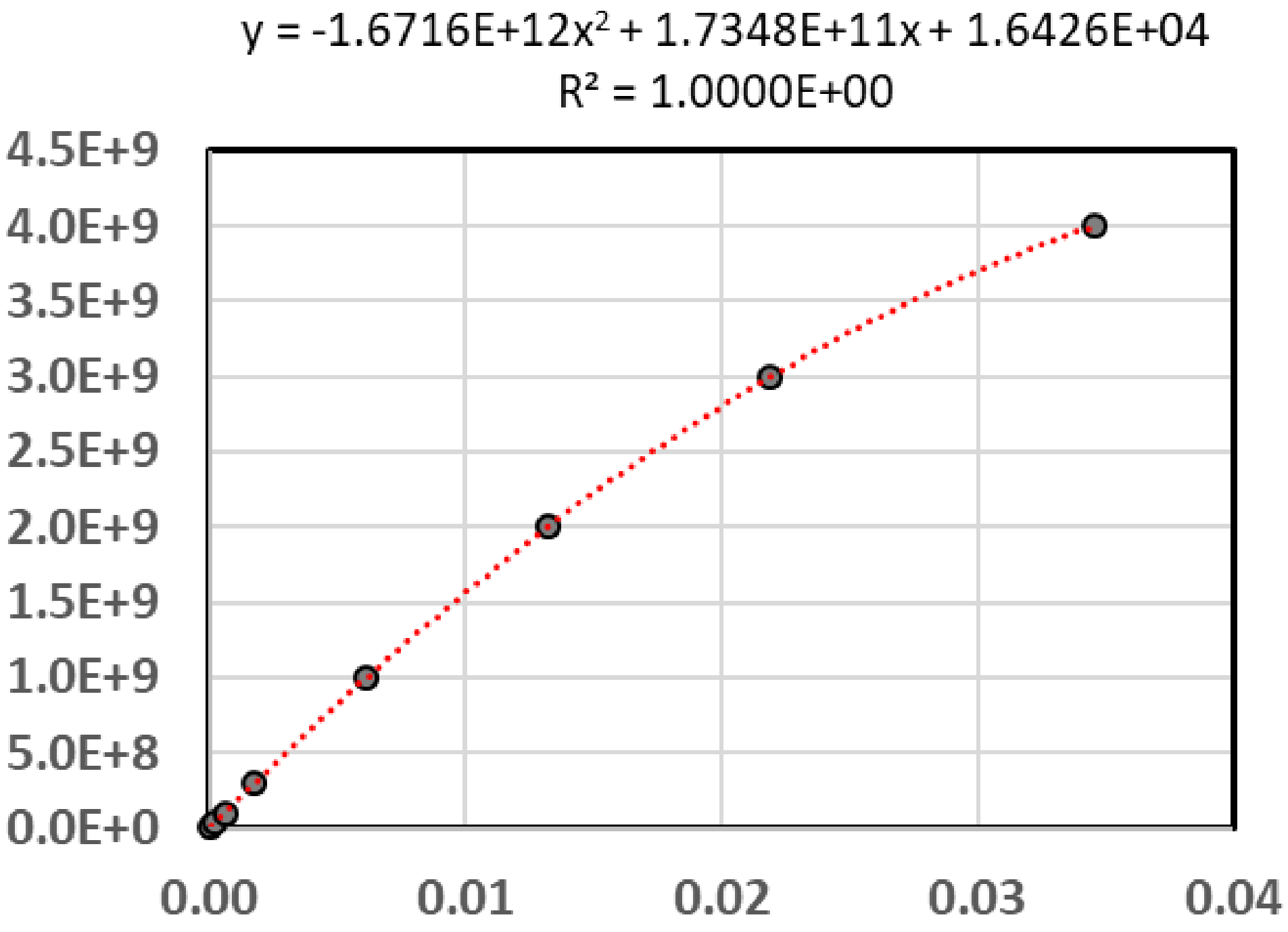

**Figures S2. Fit to 13.22 diameter crater R-values and estimated ages from Table 2.** Black points are the data points from Table 2 and the red dashed curve is the polynomial fit shown in the equation at the top of the plot.

**Caption for Table S1.** Table S1 is a comma-delimited text file (.csv) that contains a list of all craters between 7-to-25 km, their feature identification number (FID), their longitude and latitude in decimal degrees, and the unit abbreviation for the geologic unit they were assigned to.